\documentclass[journal=jpccck,manuscript=letter,layout=twocolumn]{achemso}
\usepackage{multirow}
\usepackage{chemformula} % Formula subscripts using \ch{}
\usepackage[T1]{fontenc} % Use modern font encodings
\usepackage{hyperref}
\usepackage{comment}
\usepackage[version=4]{mhchem}
\usepackage{soul}
\usepackage{lettrine}
\usepackage{siunitx}

\author{Marcelo L. Pereira Junior}
\affiliation{Department of Electrical Engineering, College of Technology, University of Bras\'ilia, 70910-900, Bras\'ilia, Federal District, Brazil}
\alsoaffiliation{Department of Materials Science and NanoEngineering, Rice University, Houston, Texas 77005, United States}

\author{Raphael M. Tromer}
\affiliation{Institute of Physics, University of Bras\'ilia, 70910-900, Bras\'ilia, Federal District, Brazil}

\author{Luiz A. Ribeiro Junior}
\affiliation{Institute of Physics, University of Bras\'ilia, 70910-900, Bras\'ilia, Federal District, Brazil}
\alsoaffiliation{Computational Materials Laboratory, University of Bras\'ilia, 70910-900, Bras\'ilia, Federal District, Brazil}

\author{Douglas S. Galv\~ao}
\affiliation{Department of Applied Physics, State University of Campinas, 13083-970, Campinas, S\~ao Paulo, Brazil}
\email{galvao@ifi.unicamp.br}

\title[]
  {
  Endohedral Derivatives of the Recently Synthesized Two-Dimensional Fullerene Networks: Electronic and Optical Insights from First-Principles Calculations
  }


\begin{document}

\begin{abstract}
\noindent The quasi-hexagonal phase of the two-dimensional fullerene network (qHPC$_{60}$), recently synthesized, has emerged as a stable carbon-based material with distinct structural and electronic features. In this work, we employed density functional theory (DFT) calculations to investigate the electronic and optical properties of its endohedral derivatives. The encapsulation of nitrogen, cerium, and strontium atoms inside fullerene cages was systematically analyzed at different concentrations. Our results show that encapsulation preserves the semiconducting backbone of pristine qHPC$_{60}$ while introducing localized electronic states that alter the bandgap and enable new transition channels. Nitrogen encapsulation produces intragap states with potential relevance for discrete optical emission, whereas cerium and strontium generate intraband states near the conduction edge. These modifications induce a red shift of the absorption onset into the visible spectrum, accompanied by enhanced refractive and absorptive responses. The robustness of the electronic structure under reduced concentrations indicates that the fully encapsulated limit adequately represents the system. Overall, the findings highlight impurity-endowed qHPC$_{60}$ as a promising platform for optoelectronic and light-harvesting applications.
\end{abstract}

\section{Introduction}

Carbon-based two-dimensional (2D) materials have been extensively investigated due to their mechanical, optical, and electronic properties \cite{gao2017flexible,zhang2019art,glavin2020emerging}. In general, they exhibit high mechanical and thermal stability, good electron mobility, and optical transparency, which make them promising candidates for the development of optoelectronic devices \cite{fan2021biphenylene,PhysRevB.70.085417,Alsayoud2018,wang2015phagraphene,wang2018popgraphene,zhuo2020me,karaush2014dft,toh2020synthesis}.  

Among the already synthesized structures, graphene is one of the most important (together with fullerenes and nanotubes), combining high electrical conductivity, mechanical robustness, and broad applicability in different technological contexts \cite{novoselov2004electric,novoselov2011nobel}. Its discovery and impact stimulated the search for other 2D carbon allotropes with complementary properties and the potential to overcome some of graphene's limitations. Within this effort, it is worth mentioning the synthesis of biphenylene network (BPN) \cite{fan2021biphenylene,pereira2022mechanical}, $\gamma$-graphyne ($\gamma$-GY) \cite{desyatkin2022scalable, RayPNAS2025,li2018synthesis,li2020review}, and monolayer amorphous carbon (MAC) \cite{toh2020synthesis,junior2021reactive,felix2020mechanical}. More recently, C$_{60}$ fullerene monolayers (2DC$_{60}$) \cite{hou2022synthesis,tromer2022dft,junior2022thermal,mortazavi2022low,peng2023stability,loh2022old,paupitz2026concise,wang2025comprehensive} have attracted much interest because they present a topology that combines closed and open cage-like motifs, sometimes described as 2.5-dimensional materials. This configuration is particularly suitable for applications such as encapsulation, gas capture, lithium storage, catalysis, and lubrication \cite{krachmalnicoff2016dipolar,dong2015b40,tan2022fullerene,shan2017functionalized,puente2021fullerenes,maroto2014chiral,rapoport2003fullerene,rapoport1999inorganic}.  

Recent studies have investigated the fundamental properties of 2DC$_{60}$. These include anisotropic thermal expansion associated with different types of intermolecular bonding \cite{shaikh2025negative}, the effect of strain engineering on thermoelectric performance \cite{wang2024strain}, interfacial thermal conductance in graphene/qHP C$_{60}$ heterostructures \cite{wang2025phonon}, charge transport mediated by large polarons and strongly influenced by lattice anisotropy \cite{cassiano2024large}, and hydrogen storage capacity \cite{ren2023enhanced}. These investigations confirm that 2DC$_{60}$ represents a stable and versatile platform with potential for a wide range of technological applications.  

Endohedral fullerenes (endofullerenes) form another very important class of fullerene derivatives, in which atoms or small molecules are encapsulated inside the C$_{60}$  or larger fullerenes \cite{gonzalez2017investigation,lu2012current,wu2024stability,zhao2024two}. Encapsulation can significantly alter the electronic, optical, and magnetic properties of the system \cite{levitt2013spectroscopy}. Several synthesis routes have been reported for endofullerenes, including chemical reactions \cite{murata2006synthesis}, high-pressure techniques \cite{pei2019recent}, and solid-state methods \cite{hoffman2021solid}.  

%Although significant progress has been made in the synthesis and characterization of 2DC$_{60}$ and in the study of endofullerenes in isolated cages and van der Waals solids, the investigation of endofullerenes in two-dimensional C$_{60}$ networks remains largely unexplored. Understanding how encapsulated species influence the electronic and optical properties of these 2D architectures is crucial for advancing their applications in energy-related, catalytic, and optoelectronic fields.  

Recently, Baldoví \textit{et al.} reported the theoretical realization of graphendofullerenes, 2D networks composed of covalently bonded endohedral C$_{80}$ units encapsulating V$_3$N clusters \cite{lopez2025graphendofullerene}. Their results revealed an interesting magnetic response arising from the interaction between the encapsulated clusters and the extended C$_{80}$ framework, demonstrating that the host–guest coupling can significantly modify the collective electronic behavior of 2D carbon systems. Concerning C$_{60}$ units, which constitute the fundamental building blocks of recently synthesized 2D fullerene materials \cite{hou2022synthesis}, Zhao and collaborators investigated, through Density Functional Theory (DFT) calculations, endohedral qHPC$_{60}$ systems modified by C$_{60}$ units encapsulating uranium and thorium dimers (U$_2$ and Th$_2$). They identified the formation of single bonds between the encapsulated metal atoms and magnetic properties induced by the fullerene confinement \cite{zhao2025design}. Nevertheless, although significant progress has been achieved in the synthesis and characterization of the 2DC$_{60}$ system, as well as in the study of endofullerenes in isolated cages and van der Waals solids, the investigation of endofullerenes in two-dimensional C$_{60}$ networks remains largely unexplored. Understanding how encapsulated species influence the electronic and optical properties of these 2D architectures is crucial for advancing their applications in energy-related, catalytic, and optoelectronic fields.

In this work, we have carried out a comprehensive DFT-based investigation on the electronic and optical properties of endofullerene derivatives based on the closely packed quasi-hexagonal C$_{60}$ monolayer (EqHPC$_{60}$) phase. We have considered the encapsulation of nitrogen (N), strontium (Sr), and cesium (Ce) atoms, which have previously been studied in van der Waals crystals formed by C$_{60}$ \cite{Martinez_Flores2022,Basiuk2011,renzler2017positively,rose1993endohedral,gao2013nitrogen}. The results reveal semiconducting electronic band gaps with localized states for N-encapsulation, as well as a metallic behavior for Ce and Sr.

\section{Methodology}

\begin{figure*}[!htb]
\centering
\includegraphics[width=0.75\linewidth]{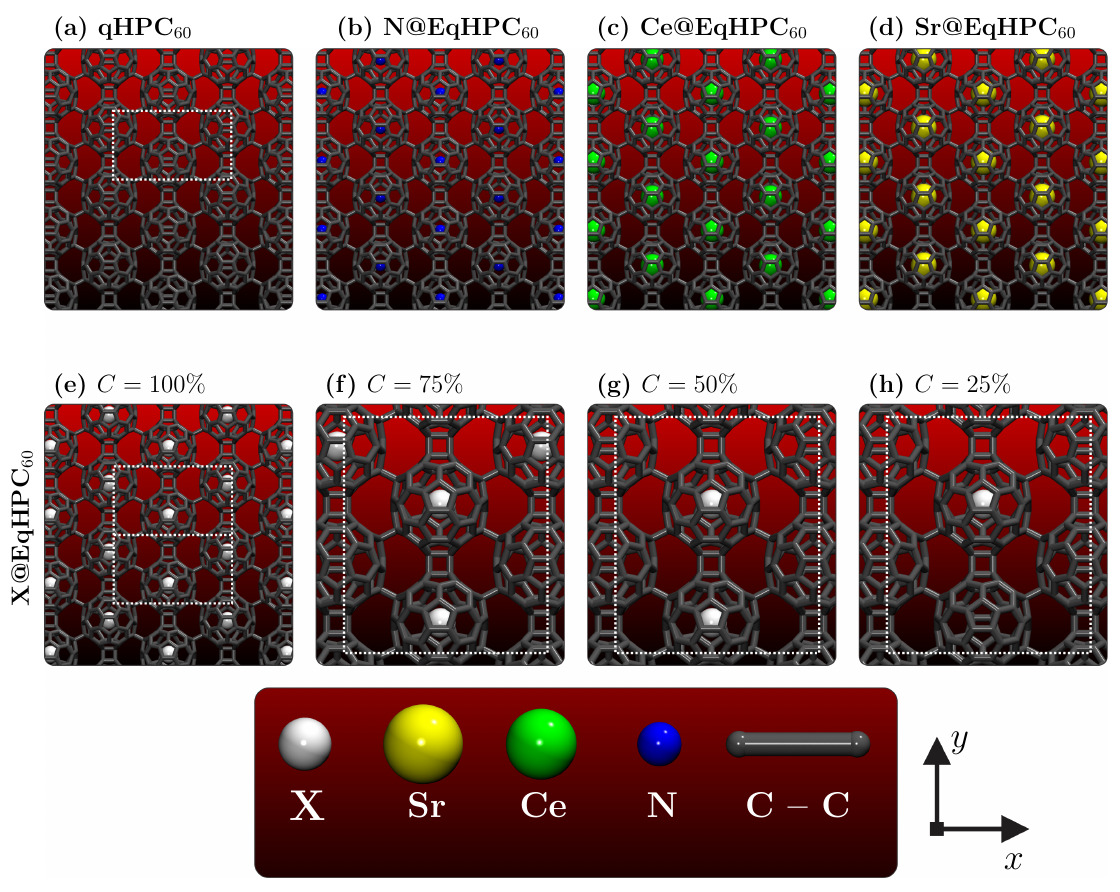}
\caption{Structures of qHPC$_{60}$ and endohedral variations. Pristine qHPC$_{60}$ (a), encapsulation with N (b), Ce (c), and Sr (d), and encapsulation fractions of 100\% (e), 75\% (f), 50\% (g), and 25\% (h). Here, X denotes a generic encapsulated atom (N, Ce, or Sr).}
\label{fig1}
\end{figure*}

As mentioned above, we have considered the quasi-hexagonal phase of C$_{60}$ (qHPC$_{60}$), which has been identified as the most stable configuration for 2DC$_{60}$ \cite{tromer2022dft}. This phase is described by a rectangular unit cell containing 120 carbon atoms, corresponding to two C$_{60}$ molecules, as illustrated in Fig.~\ref{fig1}(a). From this pristine cell, endohedral systems were generated by placing one atom of N, Ce, and Sr inside each C$_{60}$ cage, Figs.~\ref{fig1}(b)-(d). When both cages of the unit cell were filled, the fully encapsulated configuration with 100\% concentration was obtained, as schematically depicted in Fig.~\ref{fig1}(e). To explore other concentrations, a $1\times2\times1$ supercell, containing four C$_{60}$ molecules, was created. 

We have considered five cases, all encapsulated fullerenes, removing one, two, and three atomic species: 75\% (Fig.~\ref{fig1}(f)), 50\% (Fig.~\ref{fig1}(g)), 25\% (Fig.~\ref{fig1}(h)) concentrations, respectively.  Fig.~\ref{fig1} summarizes the setup of the investigated systems, from the pristine structure to different degrees of filling of the C$_{60}$ cages.

All structures were fully optimized (atomic positions and lattice parameters) prior to the electronic and optical analyses. Structural relaxations and property calculations were performed within DFT, including van der Waals corrections \cite{vdw1,vdw2,vdw3}, as implemented in the SIESTA code \cite{Soler2002}. Localized basis functions with a single-zeta (SZ) set were employed. The obtained lattice parameters, electronic band structures, and optical properties were found to be consistent with previous investigations of qHPC$_{60}$ \cite{tromer2022dft}, validating the methodology.

For lanthanide atoms such as cerium, the presence of 4$f$ orbitals leads to a large number of degenerate states near the Fermi level, which sometimes makes the convergence of the self-consistent field (SCF) cycle difficult. To address this difficulty, thermal smearing was applied \cite{Basiuk2020}. Exchange-correlation effects were described within the generalized gradient approximation (GGA) using the Perdew-Burke-Ernzerhof (PBE) functional \cite{perdew1996generalized,ernzerhof1999assessment}. Core electrons were treated with norm-conserving Troullier-Martins pseudopotentials \cite{troullier1991efficient}.

Convergence tests were performed for the energy cutoff, $k$-point mesh, and vacuum spacing. Based on these tests, an energy cutoff of 200 Ry, a Monkhorst-Pack $5\times5\times1$ $k$-point grid, and a vacuum spacing of 30 \r{A} along the $z$-direction were adopted to prevent spurious interactions between periodic (mirror) images. Atomic positions and in-plane lattice vectors ($x$ and $y$) were fully relaxed, while the out-of-plane vector ($z$) was kept fixed. The convergence criterion for maximum forces in each atom was set to 0.05 eV/\r{A}.

The electronic band structure was calculated along the high-symmetry path of the rectangular Brillouin zone defined as $\Gamma = (0.0,0.0)$ to $X = (0.5,0.0)$, $U = (0.5,0.5)$, $Y = (0.0,0.5)$, and back to $\Gamma = (0.0,0.0)$. An external electric field of 1.0 V/\r{A} was applied along three polarization planes to estimate the optical coefficients \cite{Fadaie2016}. The optical coefficients were extracted from the complex dielectric function, where the imaginary part was obtained using Fermi’s golden rule \cite{Fermi} and the real part from the Kramers-Kronig relation \cite{kronig}. From these functions, we calculated the absorption coefficient $\alpha$ according to:  
\begin{equation}
\alpha (\omega)=\sqrt{2}\omega\bigg[(\epsilon_1^2(\omega)+\epsilon_2^2(\omega))^{1/2}-\epsilon_1(\omega)\bigg ]^{1/2},
\end{equation}
the refractive index $\eta$ using: 
\begin{equation}
\eta(\omega)= \frac{1}{\sqrt{2}} \bigg [(\epsilon_1^2(\omega)+\epsilon_2^2(\omega))^{1/2}+\epsilon_1(\omega)\bigg ]^{2},
\end{equation}
and the reflectivity $R$ given by:  
\begin{equation}
R(\omega)=\bigg [\frac{(\epsilon_1(\omega)+i\epsilon_2(\omega))^{1/2}-1}{(\epsilon_1(\omega)+i\epsilon_2(\omega))^{1/2}+1}\bigg ]^2.
\end{equation}

\section{Results}

The initial analysis focuses on the structural and energetic aspects of the investigated systems. Table \ref{tab1} summarizes the lattice parameters, cohesive energies ($E_\mathrm{cohe}$), formation energies ($E_\mathrm{form}$), and the charge accumulated in the C$_{60}$ units for the pristine qHPC$_{60}$ and for the endohedral configurations containing N, Ce, and Sr, hereafter denoted as N@EqHPC$_{60}$, Ce@EqHPC$_{60}$, and Sr@EqHPC$_{60}$.

\begin{table*}[!htb]
    \begin{tabular}{|c|c|c|c|c|c|}
    \hline
    Structure & $E_\mathrm{cohe}$ (eV/atom) & $E_\mathrm{form}$ (meV/atom) & $Q_\mathrm{qHPC_{60}}$ (e) & $a$ (\r{A}) & $b$ (\r{A}) \\
    \hline
    qHPC$_{60}$ & $-7.10$ & --- & --- & 16.04 & 9.26 \\
    \hline
    N@EqHPC$_{60}$ & $-7.02$ & $-40.79$ & +0.001 & 16.35 & 9.47 \\
    \hline
    Ce@EqHPC$_{60}$ & $-7.17$ & $-178.13$ & +0.033 & 16.36 & 9.48 \\
    \hline
    Sr@EqHPC$_{60}$ & $-7.13$ & $-145.18$ & +0.032 & 16.35 & 9.47 \\
    \hline
    \end{tabular}
    \caption{Structural parameters including cohesive energy ($E_\mathrm{cohe}$), formation energy ($E_\mathrm{form}$), charge at the C$_{60}$ units, and lattice vectors ($a$ and $b$).}
    \label{tab1}
\end{table*}

The $E_\mathrm{cohe}$ was obtained using the following expression:
\begin{equation}
    E_\mathrm{cohe} = \frac{E_\mathrm{X@EqHPC_{60}} - N_\mathrm{C}E_\mathrm{C} - N_\mathrm{X}E_\mathrm{X}}{N_\mathrm{C} + N_\mathrm{X}},
\end{equation}
where $E_\mathrm{X@EqHPC_{60}}$, $E_\mathrm{C}$, and $E_\mathrm{X}$ represent the total energy of the X@EqHPC$_{60}$ system, and the energies of isolated C and X atoms, respectively, with X = Sr, Ce, or N. $N_\mathrm{C}$ and $N_\mathrm{X}$ correspond to the number of carbon and endohedral atoms in the system. The calculated $E_\mathrm{cohe}$ values indicate that all endohedral configurations remain stable with respect to the pristine qHPC$_{60}$ structure. The $E_\mathrm{form}$ was computed as:
\begin{equation}
    E_\mathrm{form} = E_\mathrm{X@EqHPC_{60}} - E_\mathrm{qHPC_{60}} - N_\mathrm{X}E_\mathrm{X},
\end{equation}
where $E_\mathrm{qHPC_{60}}$ is the total energy of the pristine qHPC$_{60}$ system, and the remaining terms are defined as above. A consistent trend is observed between $E_\mathrm{cohe}$ and $E_\mathrm{form}$, such that more negative formation energies correspond to more stable configurations. Within this framework, Ce@EqHPC$_{60}$ emerges as the most stable case, showing the smallest values for both quantities.

\begin{figure*}[!htb]
\centering
\includegraphics[width=0.9\linewidth]{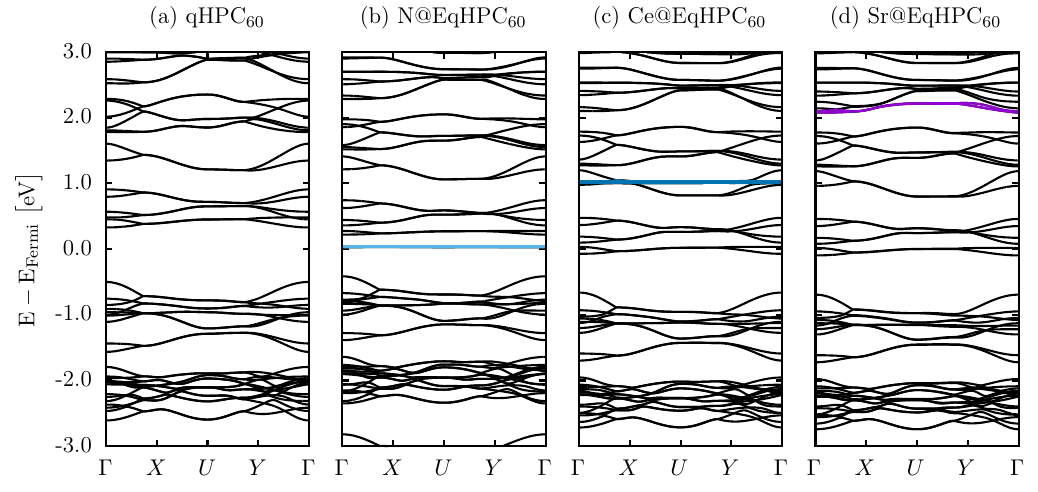}
\caption{Electronic band structures for pristine qHPC$_{60}$ (a), N@EqHPC$_{60}$ (b), Ce@EqHPC$_{60}$ (c), and Sr@EqHPC$_{60}$ (d).}
\label{fig2}
\end{figure*}

No significant transfer charges were observed. The larger values for Sr and Ce can be attributed to the differences in atomic radii and cage interactions. Sr (2.15 \r{A}) and Ce (1.81 \r{A}) are considerably larger than N (0.65 \r{A}), which brings them closer to the cage walls and increases the electrostatic coupling. Their more negative $E_\mathrm{form}$ values also reinforce the stronger interaction with the host lattice. The lattice vectors $a$ and $b$ exhibit a slight expansion compared to the pristine configuration and are basically the same for all encapsulated cases. The angle values remain essentially at 90$^\circ$, confirming that encapsulation preserves the underlying lattice symmetry.

The electronic band structures for the pristine and endohedral systems are shown in Figure \ref{fig2}, which includes qHPC$_{60}$ as well as the encapsulated configurations N@EqHPC$_{60}$, Ce@EqHPC$_{60}$, and Sr@EqHPC$_{60}$.

As a reference, pristine qHPC$_{60}$ displays a direct electronic bandgap at the $\Gamma$ point of approximately 0.86 eV, in agreement with previous reports, with the valence and conduction bands dominated by C$_{2p}$ orbitals. This configuration serves as the baseline for evaluating the modifications introduced by encapsulation.

For N@EqHPC$_{60}$ (see Fig. \ref{fig2}(b)), a localized level emerges within the gap, associated with N$_{2p}$ orbitals. This intragap state reduces the effective bandgap to approximately 0,46 eV, introduces additional recombination channels, and indicates a localized regime compatible with discrete emission. There is a conceptual parallel with wide-bandgap semiconductors such as hBN, where localized states have been identified as the origin of single-photon emission (SPE) at room temperature \cite{chatterjee2025room}. Although confirmation of such behavior in qHPC$_{60}$ requires further investigation, this result suggests potential applications in optical and quantum information.

Figure \ref{fig2}(c) shows the case of Ce@EqHPC$_{60}$, which exhibits a nearly flat state in the conduction region, located approximately 1 eV above the Fermi level, predominantly of Ce$_{4f}$ character, with noticeable hybridization with C$_{2p}$ states. This feature reflects the charge transfer from the encapsulated atom to the carbon cage and is consistent with the more negative formation energy observed for the Ce-containing system. As a consequence, the material displays a metallic character, with the Fermi level crossing the conduction band. These properties indicate a strong electronic coupling between the encapsulated atom and the fullerene cage, which may influence charge transport mechanisms and enable additional optical transitions near the conduction band edge.

A similar behavior is observed in Figure \ref{fig2}(d) for Sr@EqHPC$_{60}$, where a nearly flat state associated with Sr$_{5s}$ orbitals appears approximately $2.1$ eV above the Fermi level. This state exhibits slightly higher dispersion compared to that observed for the Ce@EqHPC$_{60}$ system, indicating a lower degree of localization, yet still sufficient to modify the conduction band profile. The crossing of the Fermi level with the conduction band confirms the metallic character of the system, reinforcing the role of the interaction between the encapsulated atom and the carbon cage in defining the electronic behavior. The correlation between charge transfer and formation energy, which is less pronounced in the case of N, further highlights the combined influence of the atomic radius and electrostatic interaction discussed in the structural analysis.

\begin{figure}[!t]
\centering
\includegraphics[width=\linewidth]{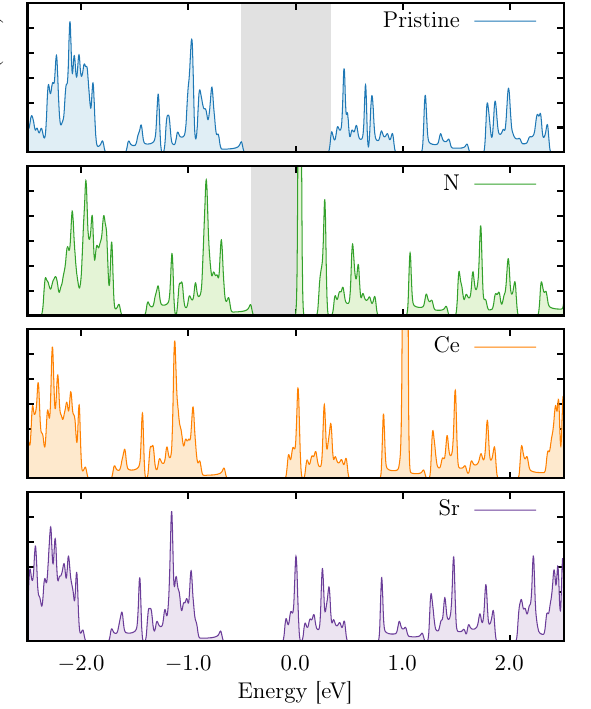}
\caption{Electronic densities of states for pristine qHPC$_{60}$ (a), N@EqHPC$_{60}$ (b), Ce@EqHPC$_{60}$ (c), and Sr@EqHPC$_{60}$ (d). Energies are aligned to $E_\mathrm{F}=0$. The shaded region indicates the bandgap.}
\label{fig3}
\end{figure}

Taken together with the band-structure analysis, Fig.~\ref{fig3} consolidates the electronic trends by displaying the Density of States (DOS) for the pristine and endohedral 2D systems, all aligned to $E_\mathrm{F}=0 eV$. The pristine system exhibits a clear gap, whereas nitrogen introduces a sharp localized level inside the gap, consistent with the flat midgap band discussed earlier. Cerium and strontium produce intense features just above $E_\mathrm{F}$ that originate from Ce$_{4f}$ and Sr$_{5s}$ states and are coherent with the weakly dispersive conduction bands near the Fermi level. These impurity-derived states are expected to alter the optical response by enabling additional transitions from the valence-band maximum (VBM) to intragap levels and from the VBM to intraband states near the conduction-band minimum (CBM). The concentration dependence of these features is examined next, followed by a quantitative analysis of the derived optical coefficients.

\begin{figure*}[!htb]
\centering
\includegraphics[width=0.75\linewidth]{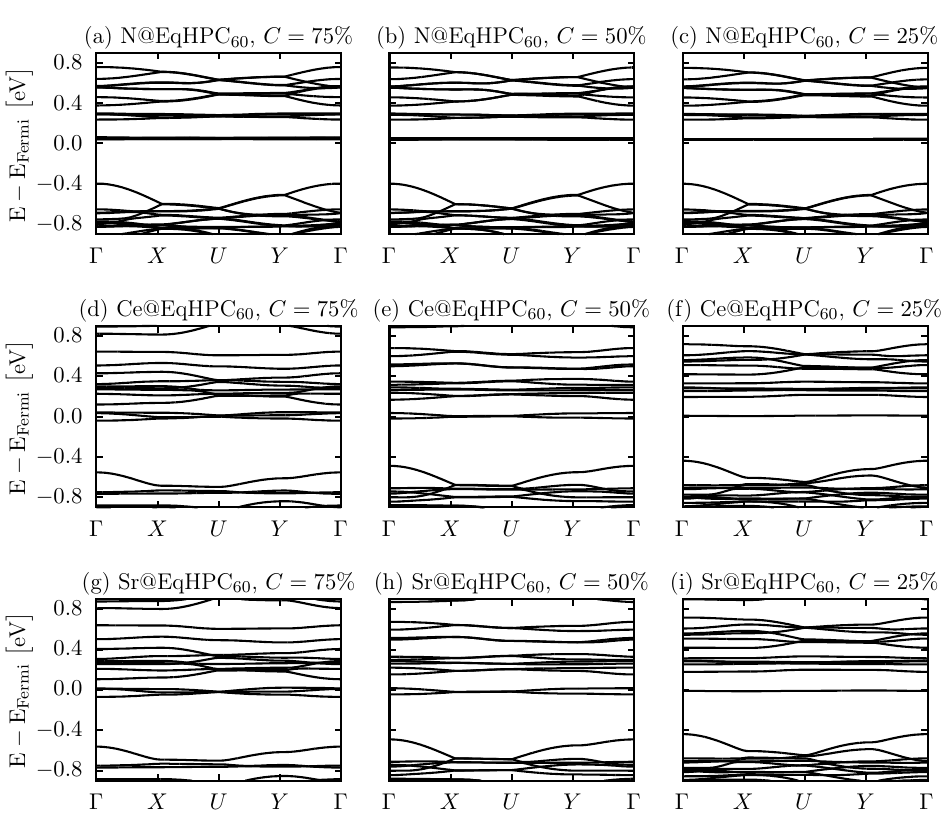}
\caption{Electronic band structures of endohedral qHPC$_{60}$ systems at different filling concentrations. Panels (a-c) correspond to N@EqHPC$_{60}$, (d-f) to Ce@EqHPC$_{60}$, and (g-i) to Sr@EqHPC$_{60}$, with filling values of 75\%, 50\%, and 25\%, respectively.}
\label{fig4}
\end{figure*}

In Figure~\ref{fig4}, the electronic band structures of N@EqHPC$_{60}$, Ce@EqHPC$_{60}$, and Sr@EqHPC$_{60}$ are presented for filling concentrations of 75\% (Fig. \ref{fig1}(f)), 50\% (Fig. \ref{fig1}(g)), and 25\% (Fig. \ref{fig1}(h)).

The results show that the main features of the EqHPC$_{60}$ band structure are preserved under partial encapsulation, confirming the structural and electronic stability of the carbon framework. The band dispersion remains nearly unchanged relative to the filled systems, with the differences limited to the intensity and energetic position of the localized states introduced by the encapsulated atoms.

For N@EqHPC$_{60}$, shown in Figure~\ref{fig4}(a–c), the systems remain semiconducting at all filling levels. The N$_{2p}$-derived intragap states appear consistently within the same energy range and maintain their separation from the band edges. The reduction in filling concentration only decreases their spectral intensity, without modifying the overall electronic character.

For Ce@EqHPC$_{60}$, shown in Figure~\ref{fig4}(d–f), nearly flat Ce$_{4f}$ states persist near the Fermi level. At 75\% and 50\% fillings, these states overlap with the conduction band, producing metallic behavior comparable to that of the fully encapsulated case. At 25\%, the system becomes semiconducting, with mid-gap states into the conduction band and a bandgap of approximately 0.44 eV.

A similar trend is observed for Sr@EqHPC$_{60}$ in Figure~\ref{fig4}(g–i). For 75\% and 50\% fillings, Sr$_{5s}$ states intersect the valence band, resulting in metallic behavior. At 25\%, the system exhibits an electronic bandgap of approximately 0.18 eV, with mid-gap states located in the valence band. The decrease in Sr concentration weakens hybridization between Sr and the carbon framework, thereby increasing electronic localization.

\begin{figure*}[!htb]
\centering
\includegraphics[width=0.9\linewidth]{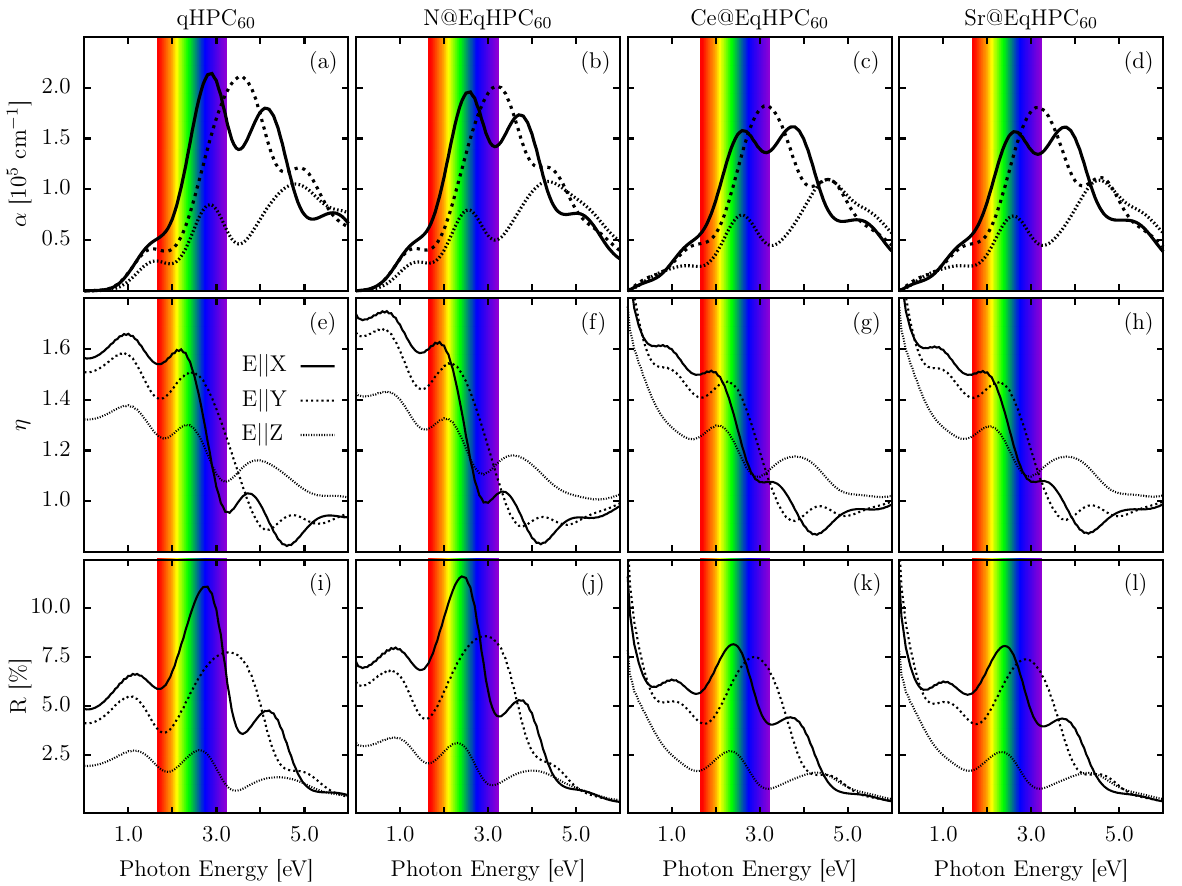}
\caption{Simulated absorption spectra ($\alpha$), refractive index ($\eta$), and reflectivity ($R$) as a function of photon energy for pristine qHPC$_{60}$ (a,e,i), N@EqHPC$_{60}$ (b,f,j), Ce@EqHPC$_{60}$ (c,g,k), and Sr@EqHPC$_{60}$ (d,h,l). Solid, dashed, and dotted lines correspond to E$\parallel$X, E$\parallel$Y, and E$\parallel$Z polarization directions, respectively. Colored bands indicate the visible spectrum range.}
\label{fig5}
\end{figure*}

The optical response of pristine qHPC$_{60}$ and their investigated endo-structures is summarized in Figure \ref{fig5}, which presents the simulated absorption spectra ($\alpha$), refractive index ($\eta$), and the reflectivity ($R$) as a function of photon energy for the three light polarizations directions E$\parallel$X, E$\parallel$Y, and E$\parallel$Z.

For the pristine case, shown in Figure \ref{fig5}(a, e, and i), the optical response is markedly anisotropic across the three polarization directions. The first absorption peak is located around 2.6 eV for E$\parallel$X and E$\parallel$Z, within the blue region of the visible spectrum. For E$\parallel$Y, it is shifted to the ultraviolet region near 3.5 eV. This transition is associated with the excitation between the highest occupied crystalline orbital (HOCO) and the lowest unoccupied crystalline orbital (LUCO). Subsequent peaks lie in the ultraviolet region, consistent with typical carbon-based optical responses \cite{xu2022rational}.

For the endo-filled systems, illustrated in Figure \ref{fig5}(b-d), the first absorption peaks are red-shifted compared to the pristine case, and all fall within the visible range. For E$\parallel$X and E$\parallel$Z, the initial transitions occur around 2.2 eV, corresponding to the green region, while for E$\parallel$Y the first peak appears near 3.1 eV, within the violet region. These features are consistent with the presence of localized electronic states created by the encapsulated atoms, as identified in the band structure analysis.

The optical bandgap values were obtained from the Tauc plots derived from the calculated absorption coefficients along the three polarization directions (E$\parallel$X, E$\parallel$Y, and E$\parallel$Z). In this approach, the absorption coefficient $\alpha$ is related to the photon energy ($h\nu$) according to the expression:

\begin{equation}
(\alpha h\nu)^2 = A(h\nu - E_g^\mathrm{opt}),    
\end{equation}

where $A$ is a proportionality constant and $E_g^\mathrm{opt}$ is the optical bandgap energy. The linear portion of the plot of $(\alpha h\nu)^2$ versus $h\nu$ was extrapolated to intercept the energy axis, yielding the value of $E_g^\mathrm{opt}$. This method assumes a direct allowed electronic transition, which is appropriate for materials exhibiting strong optical absorption near the band edge.

For the pristine EqHPC$_{60}$, the extracted optical gap is approximately 0.78~eV, in close agreement with the electronic bandgap obtained from the band structure calculations. This result confirms the consistency between the electronic and optical descriptions and indicates that excitonic effects play a minor role in this system. In contrast, for the N@EqHPC$_{60}$ system, the optical bandgap decreases to about 0.48~eV, reflecting the influence of the nitrogen-induced mid-gap states. This decrease in the optical transition threshold corroborates the presence of localized states within the electronic gap and confirms that nitrogen encapsulation effectively narrows the bandgap by introducing additional electronic levels near the Fermi energy.

For the pristine EqHPC$_{60}$, the extracted optical gap is approximately 0.78~eV, in close agreement with the electronic bandgap obtained from the band structure calculations. This result confirms the consistency between the electronic and optical descriptions and indicates that excitonic effects play a minor role in this system. In contrast, for the N@EqHPC$_{60}$ system, the optical bandgap decreases to about 0.48~eV, reflecting the influence of the nitrogen-induced mid-gap states. This decrease in the optical transition threshold corroborates the presence of localized states within the electronic gap and confirms that nitrogen encapsulation effectively narrows the bandgap by introducing additional electronic levels near the Fermi energy.

\begin{figure*}[!htb]
\centering
\includegraphics[width=0.75\linewidth]{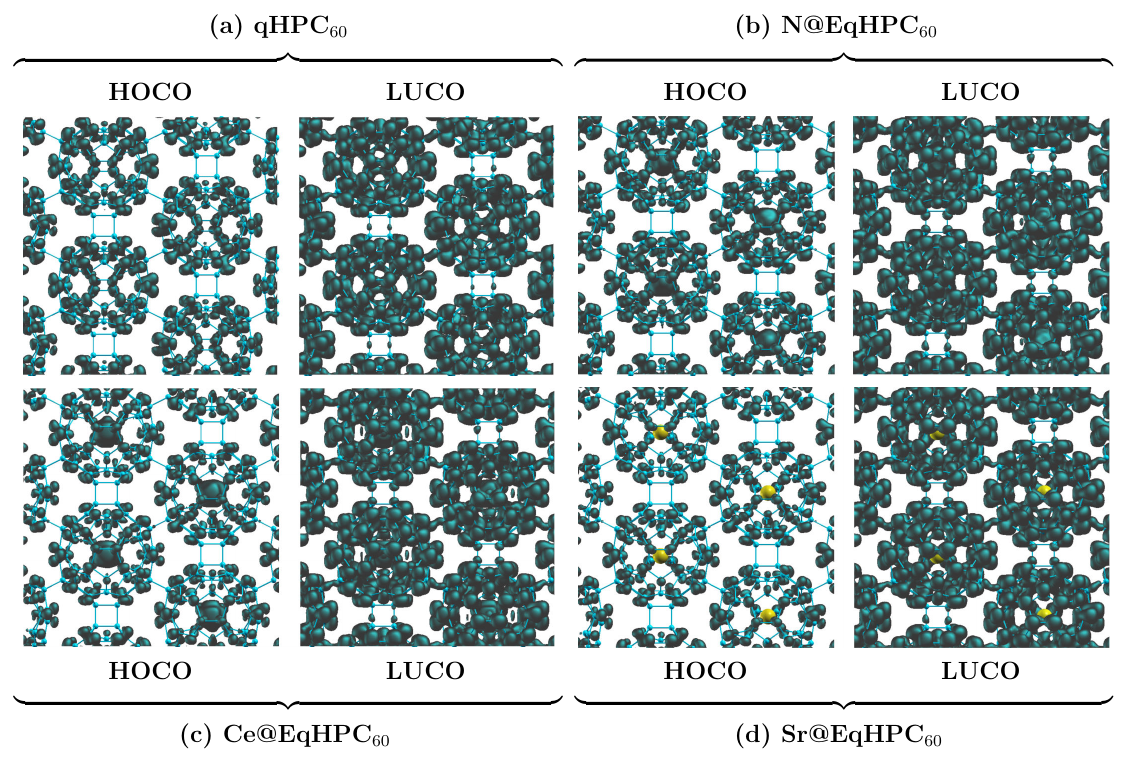}
\caption{Spatial distribution of the HOCO (left) and LUCO (right) frontier crystalline orbitals for pristine qHPC$_{60}$ (a), N@EqHPC$_{60}$ (b), Ce@EqHPC$_{60}$ (c), and Sr@EqHPC$_{60}$ (d). In the color scheme, dark green indicates regions of high electron localization, and cyan corresponds to the carbon atoms and C-C bonds.}
\label{fig6}
\end{figure*}

The reflectivity spectra, shown in Figure \ref{fig5}(i-l), also confirm the spectral shifts induced by encapsulation. In pristine qHPC$_{60}$, the reflectivity peaks occur at 2.7 eV for E$\parallel$X and E$\parallel$Z and at 3.1 eV for E$\parallel$Y. In the endo-filled systems, these peaks shift to lower energies, with maxima at approximately 2.2 eV (green region) and 2.7 eV (blue region). Within the visible spectrum, reflectivity reaches about 10\% for pristine qHPC$_{60}$ and N@EqHPC$_{60}$, while for Ce@EqHPC$_{60}$ and Sr@EqHPC$_{60}$ it decreases to around 7.5\%. These findings indicate that endohedral qHPC$_{60}$ systems exhibit low reflectivity and refractive indices greater than unity, which favors the efficient absorption of visible photons, particularly in the green, blue, and yellow regions.

In Figure \ref{fig6}, we present a schematic diagram of the spatial localization of the HOCO and LUCO frontier crystalline orbitals for all investigated systems. In general, these orbitals are distributed along the C-C bonds, forming favorable pathways for electron transport and thereby supporting high charge mobility. In the pristine system, the HOCO is mainly localized within the internal C-C bonds of the C$_{60}$ units, while the LUCO, which governs electronic mobility, is more uniformly distributed across the inter-unit bonds. A similar behavior is observed for Sr@EqHPC$_{60}$, where delocalization over the bonds remains predominant. In contrast, both HOCO and LUCO for N@EqHPC$_{60}$ and Ce@EqHPC$_{60}$ exhibit strong localization on the encapsulated atoms. These electronic states originate primarily from 2p electrons, and their localization is consistent with the reduced bandgap discussed earlier, since high charge delocalization favors a smaller separation between the frontier levels.

\section*{Conclusions}
We have performed a systematic investigation of the electronic and optical properties of endohedral fullerene networks based on the quasi-hexagonal phase of C$_{60}$. Using density functional theory, we examined pristine qHPC$_{60}$ and its endohedral variants with nitrogen, cerium, and strontium at different encapsulation concentrations. The results reveal that encapsulation maintains the semiconducting backbone of the pristine system while introducing localized states either within the bandgap or near the conduction band minimum. These states are responsible for modifying the optical absorption profile and shifting the first absorption peak into the visible region. The refractive index and reflectivity spectra further demonstrate anisotropic responses, with reduced reflectivity in the visible range, favoring efficient photon absorption. The analysis of frontier orbitals confirms that charge transport pathways are generally preserved, with impurity localization modulating the magnitude of the electronic bandgap. Notably, the concentration study demonstrates that the electronic characteristics obtained for the fully encapsulated case remain representative of partially filled lattices, supporting the robustness of the results. Taken together, these findings establish impurity-endowed qHPC$_{60}$ as a versatile candidate for applications in optoelectronics, light-harvesting, and potentially quantum information technologies.

\begin{acknowledgement}

M.L.P.J. acknowledges financial support from FAPDF (grant 00193-00001807/2023-16), CNPq (grants 444921/2024-9 and 308222/2025-3), and CAPES (grant 88887.005164/2024-00). Computational resources were provided by the National High-Performance Computing Center in São Paulo (CENAPAD-SP, UNICAMP, projects proj931 and proj960) and the High-Performance Computing Center (NACAD, Lobo Carneiro Supercomputer, UFRJ, project a22002). L.A.R.J. acknowledges the financial support from FAP-DF grants 00193.00001808/2022-71 and 00193-00001857/2023-95, FAPDF-PRONEM grant 00193.00001247/2021-20, CNPq grants 350176/2022-1, 167745/2023-9, 444111/2024-7, and PDPG-FAPDF-CAPES Centro-Oeste 00193-00000867/2024-94. We thank the Coaraci Supercomputer Center for computer time (FAPESP grant \#2019/17874-0) and the Center for Computing in Engineering and Sciences at Unicamp (FAPESP grant \#2013/08293-7). D. S. G. also acknowledges support from INEO/CNPq and FAPESP grant 2025/27044-5. The authors also thank Colin P. Nuckolls (Columbia University) for valuable discussions and suggestions that contributed to this research.
\end{acknowledgement}

%\begin{suppinfo}
%
%\end{suppinfo}

\bibliography{bibliography}

\end{document}